\documentclass[namedreferences,hyperref,optionalrh]{spr-sola}
\usepackage{graphicx}        % For eps figures, newer & more powerful
\usepackage{amssymb,amsmath,bm}        % useful mathematical symbols
\usepackage{color}           % For color text: \color command
\newcommand{\Rmnum}[1]{{\small\uppercase\expandafter{\romannumeral #1}}}
\newcommand{\Rmnums}[1]{{\scriptsize\uppercase\expandafter{\romannumeral #1}}}
\newcommand{\degr}{$\,^{\circ}$}

\newcommand{\hi}{H\,\sc{i}}
\newcommand{\heii}{He\,\sc{ii}}
\newcommand{\lya}{Ly$\alpha$}

\chardef\us=`\_

\begin{document}

\begin{frontmatter}
\title{Asymmetric Hard X-ray Radiation of Two Ribbons in a Thermal-Dominated C-Class Flare}

\author[addressref={aff1,aff2}]{\inits{G.L.}\fnm{Guanglu}~\snm{Shi}\orcid{0000-0001-7397-455X}}
\author[addressref={aff1,aff2},corref,email={lfeng@pmo.ac.cn}]{\inits{L.}\fnm{Li}~\snm{Feng}\orcid{0000-0003-4655-6939}}
\author[addressref=aff1]{\inits{J.}\fnm{Jun}~\snm{Chen}\orcid{0000-0003-3060-0480}}
\author[addressref={aff1}]{\inits{B.L.}\fnm{Beili}~\snm{Ying}\orcid{0000-0001-8402-9748}}
\author[addressref={aff1,aff2}]{\inits{S.T.}\fnm{Shuting}~\snm{Li}\orcid{0000-0003-2694-2875}}
\author[addressref={aff1,aff2}]{\inits{Q.}\fnm{Qiao}~\snm{Li}\orcid{0000-0001-7540-9335}}
\author[addressref={aff1,aff2}]{\inits{H.}\fnm{Hui}~\snm{Li}\orcid{0000-0003-1078-3021}}
\author[addressref={aff1,aff2}]{\inits{Y.}\fnm{Ying}~\snm{Li}\orcid{0000-0002-8258-4892}}
\author[addressref={aff4}]{\inits{K.F.}\fnm{Kaifan}~\snm{Ji}\orcid{0000-0001-8950-3875}}
\author[addressref={aff1,aff2}]{\inits{Y.}\fnm{Yu}~\snm{Huang}\orcid{0000-0002-0937-7221}}
\author[addressref={aff1,aff2}]{\inits{Y.P.}\fnm{Youping}~\snm{Li}\orcid{0000-0001-5529-3769}}
\author[addressref={aff1}]{\inits{J.W.}\fnm{Jingwei}~\snm{Li}}
\author[addressref={aff1}]{\inits{J.}\fnm{Jie}~\snm{Zhao}\orcid{0000-0003-3160-4379}}
\author[addressref={aff1}]{\inits{L.}\fnm{Lei}~\snm{Lu}\orcid{0000-0002-3032-6066}}
\author[addressref={aff1}]{\inits{J.C.}\fnm{Jianchao}~\snm{Xue}\orcid{0000-0003-4829-9067}}
\author[addressref={aff1}]{\inits{P.}\fnm{Ping}~\snm{Zhang}}
\author[addressref={aff1}]{\inits{D.C.}\fnm{Dechao}~\snm{Song}\orcid{0000-0003-0057-6766}}
\author[addressref={aff1,aff2}]{\inits{Z.Y.}\fnm{Zhengyuan}~\snm{Tian}\orcid{0000-0002-2158-0249}}
\author[addressref={aff1,aff2}]{\inits{Y.N.}\fnm{Yingna}~\snm{Su}\orcid{0000-0001-9647-2149}}
\author[addressref={aff1}]{\inits{Q.M.}\fnm{Qingmin}~\snm{Zhang}\orcid{0000-0003-4078-2265}}
\author[addressref={aff1}]{\inits{Y.Y.}\fnm{Yunyi}~\snm{Ge}}
\author[addressref={aff1,aff2}]{\inits{J.H.}\fnm{Jiahui}~\snm{Shan}\orcid{0009-0001-4778-5162}}
\author[addressref={aff1,aff2}]{\inits{Y.}\fnm{Yue}~\snm{Zhou}\orcid{0000-0002-3341-0845}}
\author[addressref={aff1,aff2}]{\inits{J.}\fnm{Jun}~\snm{Tian}\orcid{0000-0002-1068-4835}}
\author[addressref={aff1,aff2}]{\inits{G.}\fnm{Gen}~\snm{Li}}
\author[addressref={aff1,aff2}]{\inits{X.F.}\fnm{Xiaofeng}~\snm{Liu}\orcid{0000-0002-3657-3172}}
\author[addressref={aff1,aff2}]{\inits{Z.C.}\fnm{Zhichen}~\snm{Jing}\orcid{0000-0002-8401-9301}}
\author[addressref={aff1}]{\inits{S.J.}\fnm{Shijun}~\snm{Lei}}
\author[addressref={aff1,aff3}]{\inits{W.Q.}\fnm{Weiqun}~\snm{Gan}\orcid{0000-0001-9979-4178}}
 
%   NOTE:  Just one corresponding author [corref]
\address[id=aff1]{Key Laboratory of Dark Matter and Space Astronomy, Purple Mountain Observatory, Chinese Academy of Sciences, Nanjing 210023, China}
\address[id=aff2]{School of Astronomy and Space Science, University of Science and Technology of China, Hefei 230026, China}
\address[id=aff4]{Yunnan Observatories, Chinese Academy of Sciences, Kunming 650216, China}
\address[id=aff3]{University of Chinese Academy of Sciences, Nanjing 211135, China}

\runningauthor{Shi et al.}
% \runningtitle{Asymmetric HXR radiation of two ribbons in a thermal-dominated C4.4 flare}
\runningtitle{Asymmetric HXR Radiation in a C4.4 Flare}

\begin{abstract}

The asymmetry in hard X-ray (HXR) emission at the footpoints (FPs) of flare loops is a ubiquitous feature closely associated with nonthermal electron transport. In this study, we analyze the asymmetric HXR radiation at two flare ribbons which is thermal-dominated during a long-duration C4.4 flare that occurred on March 20, 2023, combining multi-view and multi-waveband observations from the Advanced Space-based Solar Observatory (ASO-S), Solar Orbiter (SolO), and Solar Dynamics Observatory (SDO) spacecraft. We find that the {\hi} Lyman-alpha (\lya) emission captures similar features to the {\heii} $\lambda$304 emission in both light curve and spatio-temporal evolution of a pair of conjugate flare ribbons. The spectra and imaging analysis of the HXR emission, detected by Spectrometer Telescope for Imaging X-rays (STIX) in 4-18 keV, reveal that the two-ribbon flare radiation is thermal dominated by over 95\%, and the radiation source mainly concentrates on the northern ribbon, leading to an asymmetric distribution. To understand the underlying reasons for the HXR radiation asymmetry, we extrapolate the magnetic field within the active region using the nonlinear force-free field (NLFFF) model. For 78\% of the magnetic field lines starting from the northern flare ribbon, their lengths from the loop-tops (LTs) to the northern FPs are shorter than those to the southern FPs. For 62\% of the field lines, their magnetic field strengths at the southern FPs exceed those at the northern FPs. In addition, considering the larger density, $\approx1.0\times10^{10}\ {\rm cm^{-3}}$, of the low-lying flare loops ($< 32\ {\rm Mm}$), we find that the shorter path from the LT to the northern FP enables more electrons to reach the northern FP more easily after collisions with the surrounding plasma. Therefore, in this thermal-dominated C-class flare, the asymmetric location of the flare LT relative to its two FPs plays a dominant role in the HXR radiation asymmetry, while such asymmetry is also slightly influenced by the magnetic mirror effect resulting in larger HXR radiation at the FPs with weaker magnetic strength. Our study enriches the understanding of particle transport processes during solar flares.

\end{abstract}
\end{frontmatter}

%-------------------------------------------------

\section{Introduction} \label{S-Introduction} 

Solar flares are known to accelerate particles and heat plasma, producing hard X-ray (HXR) bursts and gamma-ray bursts \citep{Hudson1995ARA,Holman2011SSRv}. To investigate the underlying physical processes of plasma heating, particle acceleration, and magnetic reconnection, spectra and imaging analysis of the HXR emission during flares serve as crucial methods \citep{Masuda1994Natur,Liu2013ApJ,Su2013NatPh,Krucker2014ApJ,Li2022ApJ}. The enhancement of X-ray fluxes mainly originates from the thermal emission that heats the plasma to a high temperature and nonthermal bremsstrahlung emission associated with the energy release of energetic particles \citep{Brown1971SoPh,Fletcher2011SSRv,Lin2011SSRv,Raymond2012SSRv}. HXR sources are often observed at the footpoints (FPs), loop-top (LT) and above the LT of flare loops, which have been extensively studied using instruments such as the Yohkoh \citep{Ogawara1992PASJ} and the Reuven Ramaty High-Energy Solar Spectroscopic Imager \citep[RHESSI,][]{Lin2002SoPh}.

In observations, the HXR sources located at the FPs of solar flares typically present asymmetry, which can be classified into two main types: S-type \citep[Sakao type,][]{Sakao1994PhDT} and N-type (non-Sakao type). These asymmetric phenomena are generally believed to be associated with the transport processes of nonthermal electrons in the flare loop. The S-type asymmetry of the HXR sources is often explained by the magnetic mirror effect \citep[e.g.,][]{Sakao1994PhDT,Kundu1995ApJ,Li1997ApJ,Aschwanden1999ApJ,Alexander2002SoPh,Ding2003ApJ,Siarkowski2004AA,Falewicz2006AdSpR,Silva2020SoPh,Naus2022ApJ,Qiu2023ApJ}. In this scenario, the brighter HXR FP is located in a region where the strength of the photospheric magnetic field is weak. The magnetic mirror effect plays a role in guiding and trapping nonthermal electrons, leading them to concentrate and emit more X-rays at one FP than the other. The N-type asymmetry is featured as the brighter FP of the HXR emission corresponds to the region with a stronger magnetic field, which can be interpreted as the reconnection site generated by interactions with other flare loops being closer to that FP \citep[e.g.,][]{Asai2002ApJ,Goff2004AA,Falewicz2007AA,Joshi2017ApJ}. The magnetic reconnection reduces the effect of magnetic field convergence, allowing numerous electrons to precipitate at the brighter FP.

The Advanced Space-based Solar Observatory \citep[ASO-S,][]{Gan2023SoPh}, launched in Oct. 2022, is the first comprehensive solar space observatory in China. The scientific objectives of ASO-S are focused on ``1M2B'' \citep{Gan2019RAA}, namely the magnetic fields and two types of violent eruptions on the sun -- solar flares and coronal mass ejections (CMEs). There are three coordinated payloads on board the ASO-S, the Full-disk vector MagnetoGraph \citep[FMG,][]{Deng2019RAA,Su2019RAA19}, the Lyman-alpha ({\lya}) Solar Telescope \citep[LST,][]{Li2019RAA,Chen2019RAA,Feng2019RAA}, and the Hard X-ray Imager \citep[HXI,][]{Zhang2019RAA,Su2019RAA163S}. Three payloads can be used to investigate the relationships among solar flares, CMEs, and magnetic fields. The LST payload consists of three instruments: the Solar Disk Imager (SDI), the Solar Corona Imager (SCI), and the White-light Solar Telescope (WST). These instruments enable simultaneous dual-waveband observations from the solar disk to the low corona, focusing on the studies of flare, prominence and CME eruptions.

In this work, we analyze the asymmetric HXR sources in a two-ribbon flare dominated by over 95\% thermal radiation during the very long rise phase of about two hours in a long-duration event (LDE). We further explain this phenomenon by combining the results of the nonlinear force-free field (NLFFF) extrapolation. Section~\ref{S-obs} provides an overview of the LDE observations. Section \ref{S-results} presents our analysis and results: (1) STIX HXR spectra and imaging analysis; (2) comparisons between {\hi} {\lya} and {\heii} 304 {\AA} emissions; (3) the configuration of the magnetic field extrapolated by the NLFFF model. Our conclusions and discussion are given in Section \ref{S-conclusions}.

\section{Overview of Observations} \label{S-obs}

The two-ribbon flare (SOL2023-03-20T15:34) studied in this paper was produced by the eruption of a hot channel and filament system in the active region NOAA AR 13258. This event was observed by various instruments, including the SDI instrument of the LST payload on board the ASO-S, Spectrometer Telescope for Imaging X-rays \citep[STIX,][]{Krucker2020AA} on board the Solar Orbiter \citep[SolO,][]{Muller2020AA}, the Atmospheric Imaging Assembly \citep[AIA,][]{Lemen2012SoPh} on board the Solar Dynamics Observatory \citep[SDO,][]{Pesnell2012SoPh}, the Geostationary Operational Environmental Satellite (GOES), and the Chinese H$\alpha$ Solar Explorer \citep[CHASE,][]{Li2022SCPMA}. The SolO was located at a longitude separation of 18.4{\degr} from Earth and had a heliocentric distance of 0.51 AU. Multi-view and multi-waveband observations provide detailed insights into the dynamics of the flare and enable a more comprehensive understanding of the associated particle behaviors.

This event is a GOES C4.4 flare that started at 13:30 UT and peaked at 15:34 UT on March 20, 2023, shown in Figure \ref{fig1}a, which is a long-duration event (LDE) lasted for an impressive duration of 11 hours. The curve of the time derivative (purple line) presented in Panel a is derived from the GOES data \citep{Thomas1985SoPh}. The light curves of HXR emissions at photon energies of 4-10 keV (green line) and 10-18 keV (magenta line), detected by STIX, are shown in Panel b. Panel c displays the light curves of AIA {\heii} 304 {\AA} (black line), AIA 94 {\AA} (cyan line), and SDI {\hi} {\lya} (blue line), obtained by integrating over the entire flare region. All of these light curves display a significant enhancement during the flare. This event is classified as a relatively small C-class flare, only low-energy photons (4-18 keV) were detected by STIX at a distance of about 0.51 AU. Additionally, the peak time of the SXR emission, marked by a red dashed line, is close to that of the HXR emission. It implies that the energy released during this event is probably not dominated by nonthermal electrons (further evidence is presented in Section~\ref{S-stix}). Unfortunately, due to the instrument being in the test status, the SDI only has available observations starting from 13:57 UT during this period. Despite this limitation, it is still possible to observe a good temporal correlation between the HXR emissions and the light curves of {\hi} {\lya} and {\heii} 304 {\AA}, implying that they might have similar radiation origins. In the flare rising phase, {\lya} and 304 {\AA} radiation were from flare ribbons. However, clear distinctions are evident between the HXR and 94 {\AA}.

Figure \ref{fig2} presents the flare-related structures observed by multiple instruments during this LDE. Panels a and b clearly show the associated filament and hot channel observed by CHASE H$\alpha$ and AIA 94 {\AA}, respectively. As the filament rotates and rapidly rises, it subsequently develops into a CME accompanied by the formation of an elongated current sheet (CS), a series of blobs, and supra-arcade downflows (SADs). 
%Using the 3D-gradual cylindrical shell (GCS) model, we estimate the radical speed of the CME to be 930 {\kms}. Approximately 18 blobs are detected by the white-light (WL) coronagraph, with speeds ranging from 344-876 {\kms}. 
Simultaneous observations from AIA {\heii} 304 {\AA} and SDI {\hi} {\lya} capture the spatio-temporal evolution of a pair of conjugate flare ribbons, as shown in Panels c and d, presenting a good consistency with each other. During the flare rising phase, the two ribbons rapidly separate towards the northern and southern directions, respectively. The flare ribbons can also be observed in the H$\alpha$ images. There is no significant brightening in the AIA 1600 {\AA} and 1700 {\AA} images, while brightening in these two passbands is generally believed to be associated with the motion of nonthermal electrons in the flare loops. These phenomena indicate that the flare ribbons are mainly dominated by thermal emission, and formed in the middle-to-upper chromosphere and transition region.

\section{Analysis and Results} \label{S-results}

\subsection{HXR Spectra and Imaging Analysis} \label{S-stix}

\begin{figure}
\centerline{\includegraphics[width=0.9\textwidth]{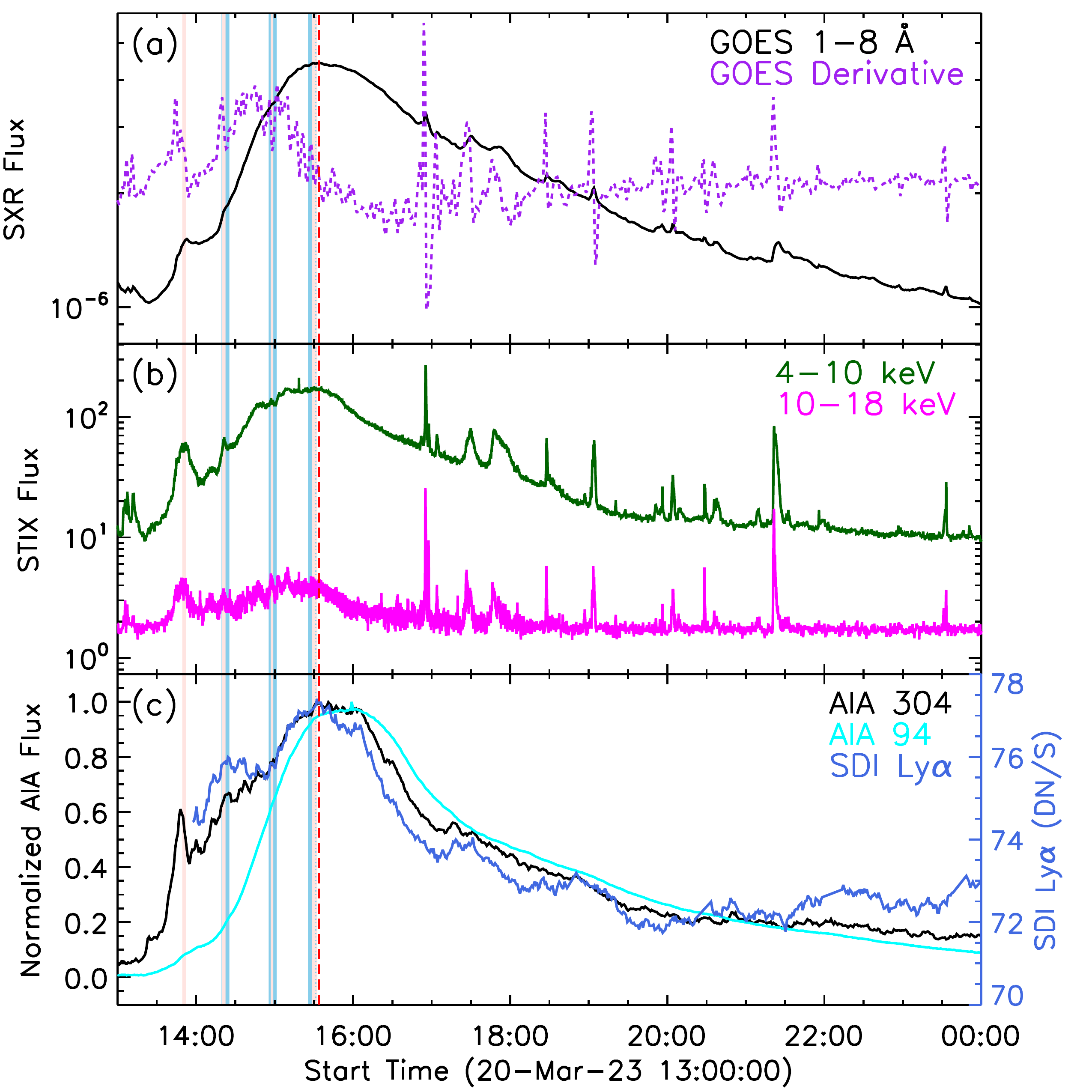}}
\caption{Light curves from multiple wavebands for the long-duration event (LDE) on March 20, 2023. (\textbf{a}) GOES SXR light curve ranging from 1.0-8.0 {\AA} along with its time derivative (purple line). (\textbf{b}) STIX HXR light curves at 4-10 keV (green line) and 10-18 keV (magenta line). (\textbf{c}) Integrated light curves of AIA {\heii} 304 {\AA} (black line), AIA 94 {\AA} (cyan line) and SDI {\hi} {\lya} (blue line) in the flare region. Pink and blue shadows represent the integration times for HXR spectra and imaging analysis shown in Figure \ref{fig3} and Figure \ref{fig4}, respectively. The red vertical line marks the SXR peak time at 15:34 UT.}
\label{fig1}
\end{figure}

\begin{figure}
\centerline{\includegraphics[width=1.0\textwidth]{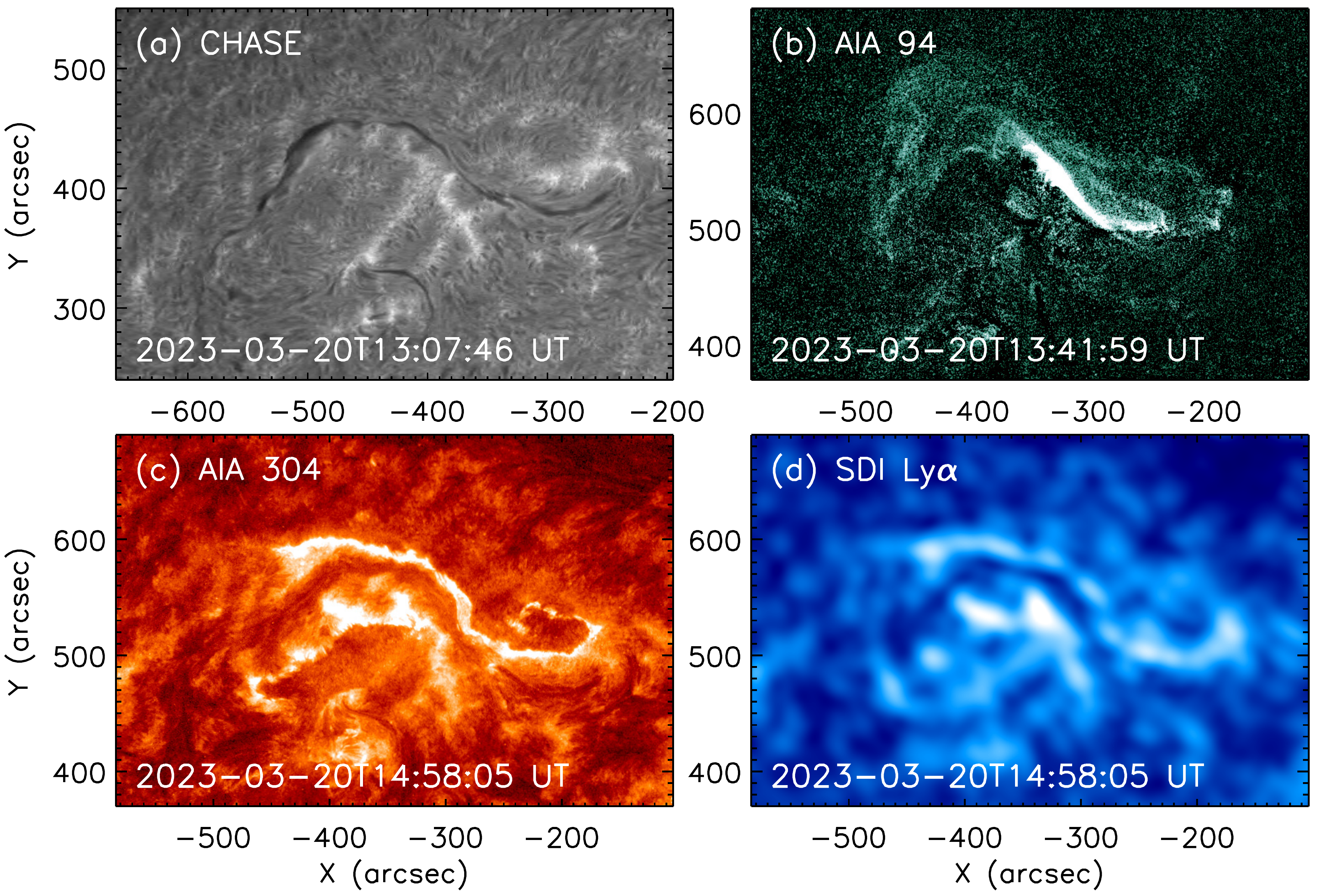}}
\caption{An overview of the observations of the LDE. (\textbf{a}) A long filament before the flare onset observed in H$\alpha$ by CHASE. (\textbf{b}) Eruption of the hot channel observed in the AIA base-difference image in the 94 {\AA} passband. (\textbf{c}) and (\textbf{d}) Separation of the two ribbons captured in AIA {\heii} 304 {\AA} (\textbf{c}) and SDI {\hi} {\lya} (\textbf{d}) towards the northern and southern directions.}
\label{fig2}
\end{figure}

\begin{figure}
\centerline{\includegraphics[width=1.0\textwidth]{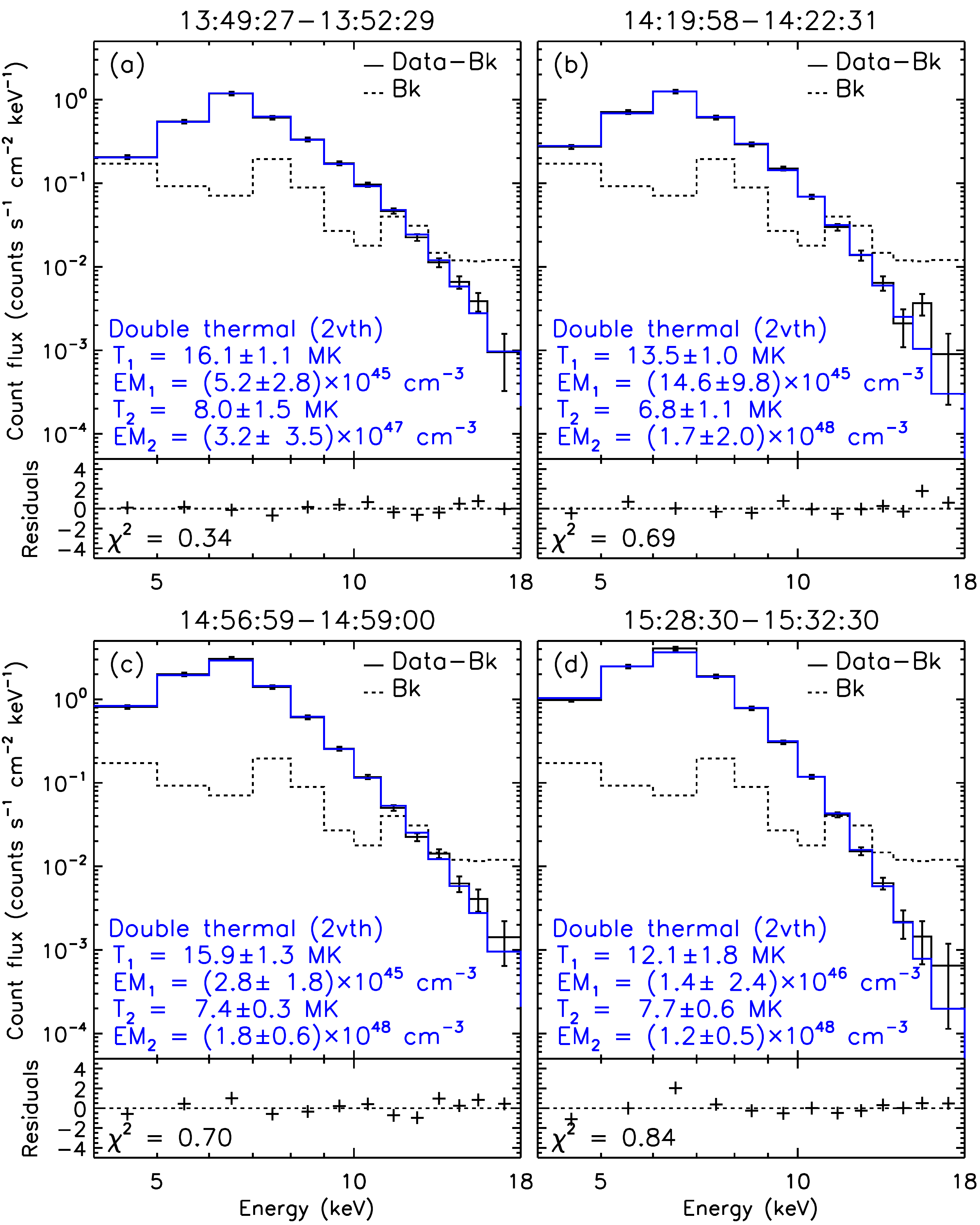}}
\caption{HXR spectral fitting for the times marked by the pink shadows in Figure \ref{fig1} using a double thermal model. The fitting parameters, including electron temperature (T), emission measure (EM), and chi-square ($\chi^2$), are displayed in each panel. Black solid lines represent the count rate of the HXR emissions detected by STIX after subtraction of the pre-flare background (dashed lines), while blue lines represent the optimal fitting curves.}
\label{fig3}
\end{figure}

\begin{figure}
\centerline{\includegraphics[width=1.0\textwidth]{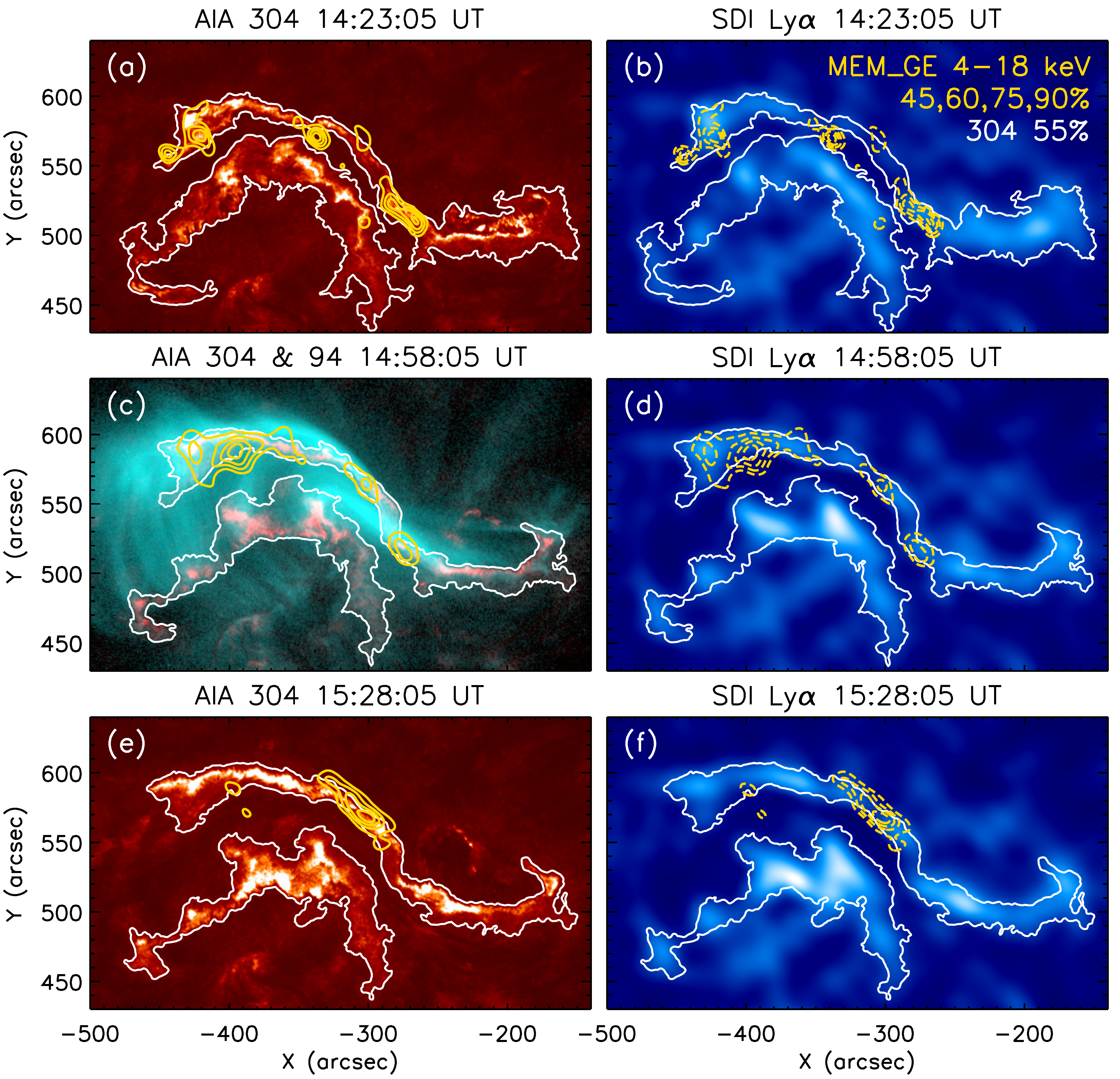}}
\caption{Evolution of two ribbons in the AIA {\heii} 304 {\AA} (left column) and SDI {\hi} {\lya} (right column) observations for the times selected by the blue shadows in Figure \ref{fig1}. Panel c presents the composite image of AIA 304 {\AA} (red) and 94 {\AA} (cyan). The overlaid gold contours represent HXR maps of STIX reconstructed by the MEM\_GE algorithm at 4-18 keV, using levels at 45\%-90\% of the maximum intensity with a step of 15\% in each panel. The white contour indicates 55\% of the maximum intensity of {\heii} 304 {\AA}.}
\label{fig4}
\end{figure}

To determine the nonthermal/thermal nature of the two-ribbon flare and clarify the origin of the HXR emission, we conduct an analysis of both the spectra and imaging of the HXR detected by STIX during the flare rising phase. Figure \ref{fig3} and Figure \ref{fig4} present the results for the selected time intervals marked by the pink and blue shadows in Figure \ref{fig1}, respectively.

First, we use a double thermal model that describes optically thin thermal bremsstrahlung radiation to fit the observed HXR spectra. Figure \ref{fig3} presents the spatially integrated count spectra (black solid lines) detected by STIX for various time intervals from 2 to 4 minutes, after subtracting the pre-flare background (black dashed lines) from 13:20 to 13:22 UT. The chosen time intervals are to ensure sufficient count rates in 4 to 18 keV. Considering the extremely long rising phase of about 2 hours for this flare, the temperature may not have significant change in 2 to 4 minutes. Blue lines in each panel represent the optimal fitting curves. Free parameters of the double thermal functions, including electron temperature (T) and emission measure (EM), are indicated in each panel. The residual distribution and the coefficient chi-square ($\chi^2$) quantitatively evaluate the level of consistency between the fitting function and the HXR spectra. We also attempted to use a standard combination of thermal and nonthermal power-law functions to fit the HXR spectra. It turns out that their $\chi^2$ values are at least twice as large as those in the double-thermal fittings. The thick-target spectral indices are about 8 to 10 during the flare rising phase, and the ratios of integrated nonthermal to thermal count fluxes are less than 5\%. Therefore, based on these two different fitting models, the spectral fitting analysis demonstrates that the two-ribbon flare radiation is thermal dominated by over 95\%. Panel a presents the spectral fitting results at 13:51 UT of the sub-peak preceding the main peak of the HXR light curves, which corresponds to the filament eruption. In Panels b-d, it can be seen that during the rising phase, the weighted mean temperature range of the flare is from 6.83 MK (14:21 UT) to 7.76 MK (15:30 UT).

Then, we reconstruct the HXR maps detected by STIX and superpose them on AIA {\heii} 304 {\AA} (left column) and SDI {\hi} {\lya} (right column) images, as shown in Figure \ref{fig4}. Panel c additionally presents the composite image combining AIA 304 {\AA} (red) and AIA 94 {\AA} (cyan) at 14:58 UT. To generate these images, visibilities are computed by subcollimators labeled from 3 to 10, with an integration time of 6 minutes centered on the observation time of AIA, and applied to the MEM\_GE algorithm \citep{Massa2020ApJ} for the thermal sources in the 4-18 keV energy range, resulting in an angular resolution of $\approx14.6''$. The co-alignment of the STIX images was done with the Full Sun Imager (FSI) 304 {\AA} images observed by the Extreme Ultraviolet Imager \citep[EUI,][]{Rochus2020} on board the SolO. Subsequently, we reprojected the STIX images from the SolO view to Earth view, assuming that all photons originate from almost the same altitude, i.e., the middle-to-upper chromosphere or transition region. The reason for selecting a 6-minute integration time is that it can obtain the HXR images more consistent with the flare ribbons observed in AIA 304 {\AA} and SDI {\lya} compared to the 4-minute images during the initial stage of the flare, and they have similar imaging results near the peak time of the GOES flux. There were no SDI {\lya} observations before 13:57 UT, and the parallel sections of the flare ribbons formed after 14:00 UT. Therefore, STIX imaging results at 13:51 UT are not presented in Figure \ref{fig4}. The gold contours in each panel indicate 45\%-90\% of the maximum HXR intensity, with a step of 15\%. The HXR radiation mainly originates from the flare FPs and presents an asymmetric distribution, primarily concentrated along the parallel section of the northern ribbon. As the flare evolves towards the peak time of the SXR flux, a weak HXR source begins to appear at the southern FP (not shown in Figure \ref{fig4}). The white contours in each panel represent 55\% of the maximum intensity of the AIA {\heii} 304 {\AA}. From comparisons, it is evident that the {\heii} 304 {\AA} and {\hi} {\lya} emissions present a good spatial consistency in the evolution of the flare FPs, consistent with the study of \citet{Li2022ApJ936}.

Note that the HXR sources are predominantly from the flare ribbon instead of the projection of the higher coronal source. From the light curves in Figure \ref{fig1}, the STIX light curve has more similarity to that of AIA 304 {\AA} than AIA 94 {\AA}. Moreover, if we compare the morphology details of the STIX HXR source with that of the flare ribbon observed in AIA 304 {\AA} and that of flare loops observed in AIA 94 {\AA}, we find that the HXR source resembles the northern ribbon more, especially at the upper-left end in Figure \ref{fig4}a and c.

The spectra and imaging analysis further support the conclusion that the HXR emission from the flare FPs is dominated by the collision of thermal electrons injected downward from magnetic reconnection with the surrounding plasma in the dense flare loop. Furthermore, both the emissions of {\heii} 304 {\AA} and {\hi} {\lya} have similar radiation features produced by the thermal process that occurred in the two ribbons of the flare.

\subsection{{\hi} {\lya} and {\heii} 304 {\AA} Emissions} \label{S-lya}

To quantitatively investigate the relationship between the radiation intensities of {\heii} 304 {\AA} and {\hi} {\lya}, we conduct pixel-by-pixel comparisons in the two-ribbon regions and calculate their Pearson correlation coefficients (CCs).

\begin{figure}
\centerline{\includegraphics[width=1.0\textwidth]{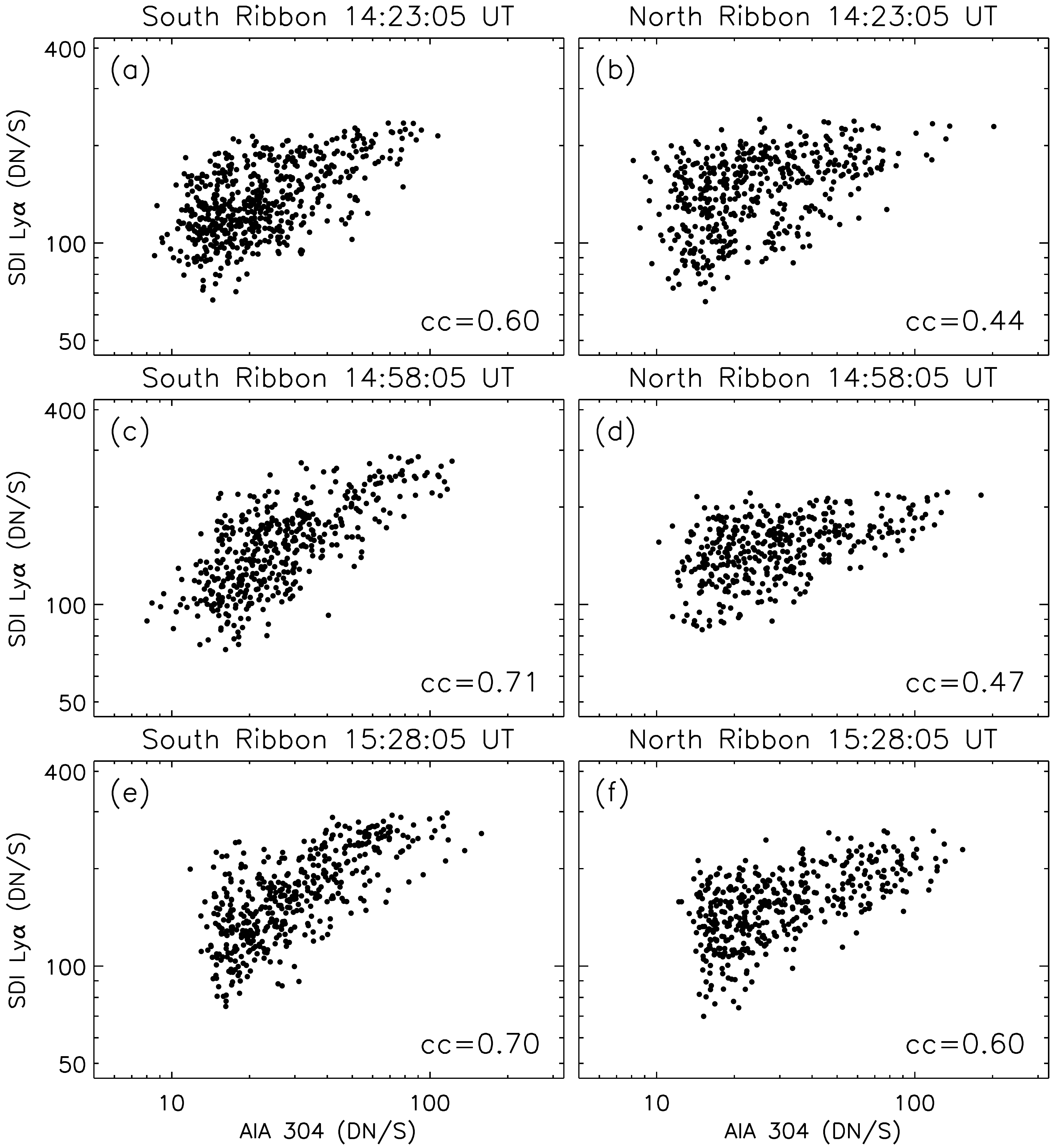}}
\caption{Scatter diagrams of the pixel-by-pixel comparisons between the AIA {\heii} 304 {\AA} (horizontal axis) and SDI {\hi} {\lya} (vertical axis) intensities within regions of the southern ribbon (left column) and the northern ribbon (right column) at different times. Correlation coefficients (CCs) are calculated in each panel.}
\label{fig5}
\end{figure}

Aligning the SDI images with the AIA using the method developed by \citet{Cai2022RAA} is crucial to ensure that the calculated radiation comes from the same flare region, enabling accurate point-to-point comparison of radiation intensities between {\heii} 304 {\AA} and {\hi} {\lya}. We then uniformly sample points within the white contour (representing 55\% of the 304 {\AA} intensity) shown in Figure \ref{fig4} and extract their corresponding radiation intensities on a pixel-by-pixel basis. Figure \ref{fig5} presents scatter diagrams of the intensity comparisons under logarithmic axes between {\heii} 304 {\AA} and {\hi} {\lya} in the southern (left column) and northern (right column) ribbon regions from 14:23 UT to 15:28 UT. One can see that the {\hi} {\lya} radiation has moderate intensity correlations with {\heii} 304 {\AA}, while the CCs of the southern ribbon (0.60-0.71) are greater than that of the northern ribbon (0.44-0.60). The calculated difference in CCs between the two ribbons may be attributed to the distinct environmental conditions in each region. Meanwhile, as the flare ribbons evolve, there is a gradual increase in the correlation between {\hi} {\lya} and {\heii} 304 {\AA} emissions, which may be closely associated with the heating process in the flare.

\subsection{NLFFF Extrapolation} \label{S-nlfff}

\begin{figure}
\centerline{\includegraphics[width=1.0\textwidth]{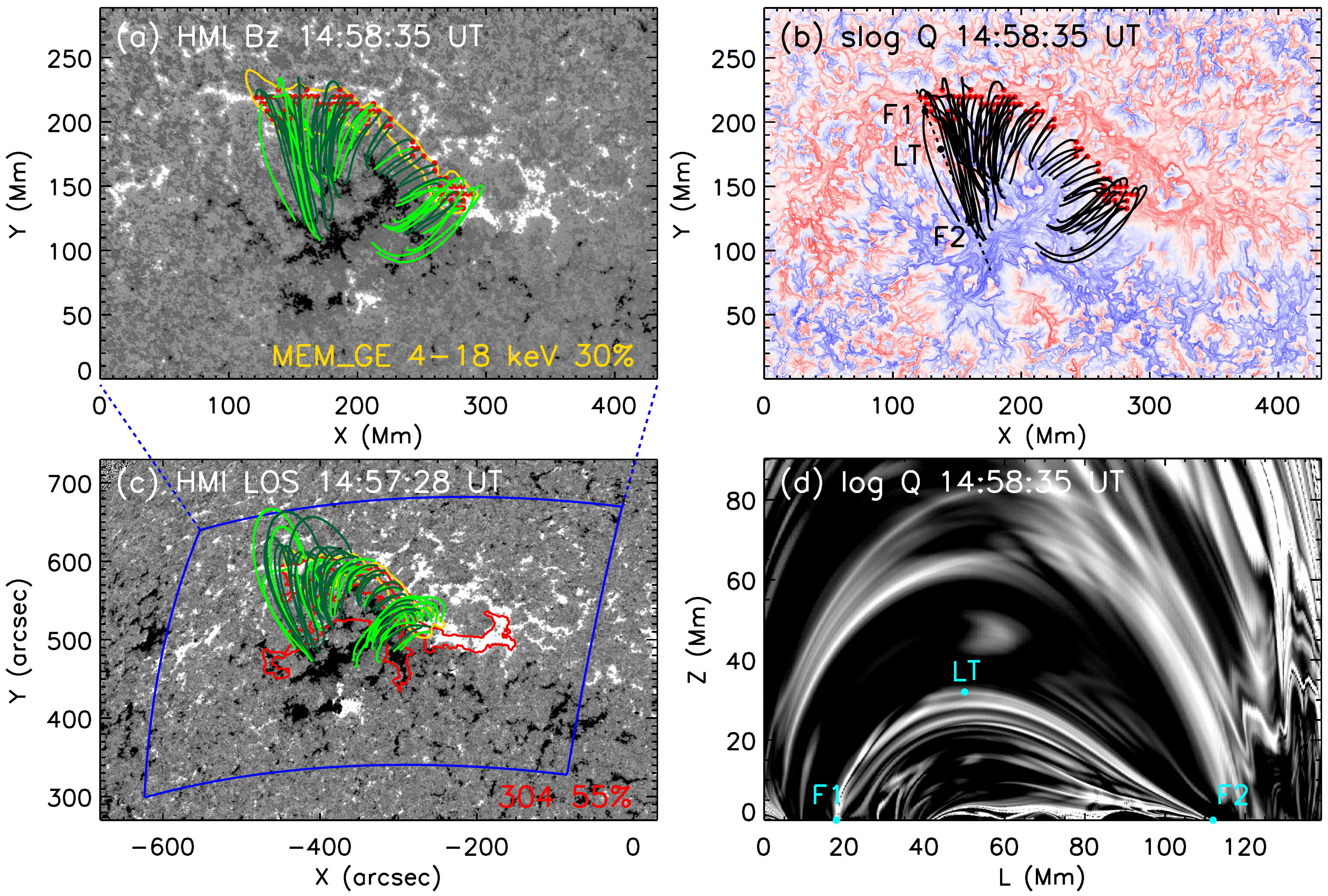}}
\caption{Magnetic configuration in the two-ribbon region. (\textbf{a}) HMI $B_z$ component of NOAA AR 13258 in the CEA coordinate applied for the NLFFF extrapolation, with a linear saturating at 300 G. Green field lines root at the red dots within the gold contour that represents the 30\% level of the maximum intensity of HXR. Dark green lines indicate a stronger magnetic field in the southern FP compared to the northern FP. (\textbf{b}) Signed log Q map, ${\rm slog}\,Q = {\rm sgn}(B_{\rm z})\cdot {\rm log}\,Q$, on the NLFFF photospheric boundary. (\textbf{c}) HMI line-of-sight (LOS) magnetogram overlaid the red contour (55\% of the maximum intensity of {\heii} 304 {\AA}) outlining the flare ribbons; the blue rectangle indicates the field-of-view (FoV) of the SHARP map in Panel a. (\textbf{d}) Log Q map, with its axes along the dashed line and Z direction in Panel b, respectively. F1 and F2 denote footpoints (FPs) of the flare loop, while LT represents the flare loop-top.}
\label{fig6}
\end{figure}

To understand the asymmetric distribution of the HXR emission and the morphology of the two ribbons, we conduct the NLFFF extrapolation based on the optimization method \citep{Wiegelmann2003NPGeo} in the flare region. The structure of the magnetic fields is presented in Figure \ref{fig6}. 

The Space-Weather HMI Active Region Patches \citep[SHARP,][]{Bobra2014SoPh} map, derived from the Helioseismic and Magnetic Imager \citep[HMI,][]{Hoeksema2014SoPh} on board the SDO, is used as the boundary condition for the NLFFF model. The 180-degree ambiguity is corrected, and the map is projected onto the cylindrical equal-area (CEA) coordinate. Figure \ref{fig6}a presents the $B_{\rm z}$ component of the SHARP map at 14:58~UT with a pixel scale of 0.36 Mm. The gold contour represents 30\% of the maximum HXR intensity projected from the Helioprojective-Cartesian (HPC) coordinate in Figure \ref{fig4}c and d to the CEA coordinate, and red dots are uniformly sampled within it as the start points for tracing magnetic field lines (green lines). The dark green lines represent magnetic field lines, where the magnetic field strengths at the southern FPs are larger than those at the northern FPs. In Panel c, the distribution of traced field lines is displayed along with the ribbon contours of AIA {\heii} 304 {\AA} (55\%, red lines) and HXR (30\%, gold lines) on the HMI Line-of-Sight (LoS) magnetogram. The blue rectangle represents the Field-of-View (FoV) of the SHARP map shown in Panel a. One can see that the configuration of the magnetic field follows a horn-like shape, featuring a convergence field rooted in the southern region with negative fluxes and a divergent distribution in the northern region with positive fluxes.

Based on the extrapolated magnetic field, we calculate the squashing factor Q \citep{Titov2002} to quantify the magnetic connectivity using the FastQSL code developed by \citet{Zhang2022ApJ}. Figure \ref{fig6}b presents the distribution of the signed log Q, ${\rm slog}\,Q = {\rm sgn}(B_{\rm z})\cdot {\rm log}\,Q$, at the photosphere. The magnetic field lines originating from the red dots marked in the high-Q region (red) in the northern region are all connected to the high-Q region (blue) in the southern region. To gain insights into the three-dimensional structure of the magnetic field, particularly the quasi-separatrix layer (QSL), we calculate the Q within the 3D box region. Panel d shows its distribution on a cross-section along the dashed line in Panel b and perpendicular to the XOY plane. From the QSL map, we can see the positions of the LT and FPs (F1, F2) of the flare loop. The LT is located at a height of $\approx32\ {\rm Mm}$ above the photosphere, while F1 and F2 are rooted in the northern and southern ribbons, respectively. One can see that the distance from the flare LT to the two FPs is asymmetric, indicating that the length from the LT to F1 is shorter than that to F2. Such magnetic field configuration suggests that for the flare loops at the eastern side, the energy release of particles is located closer to F1.

\begin{figure}
\centerline{\includegraphics[width=1.0\textwidth]{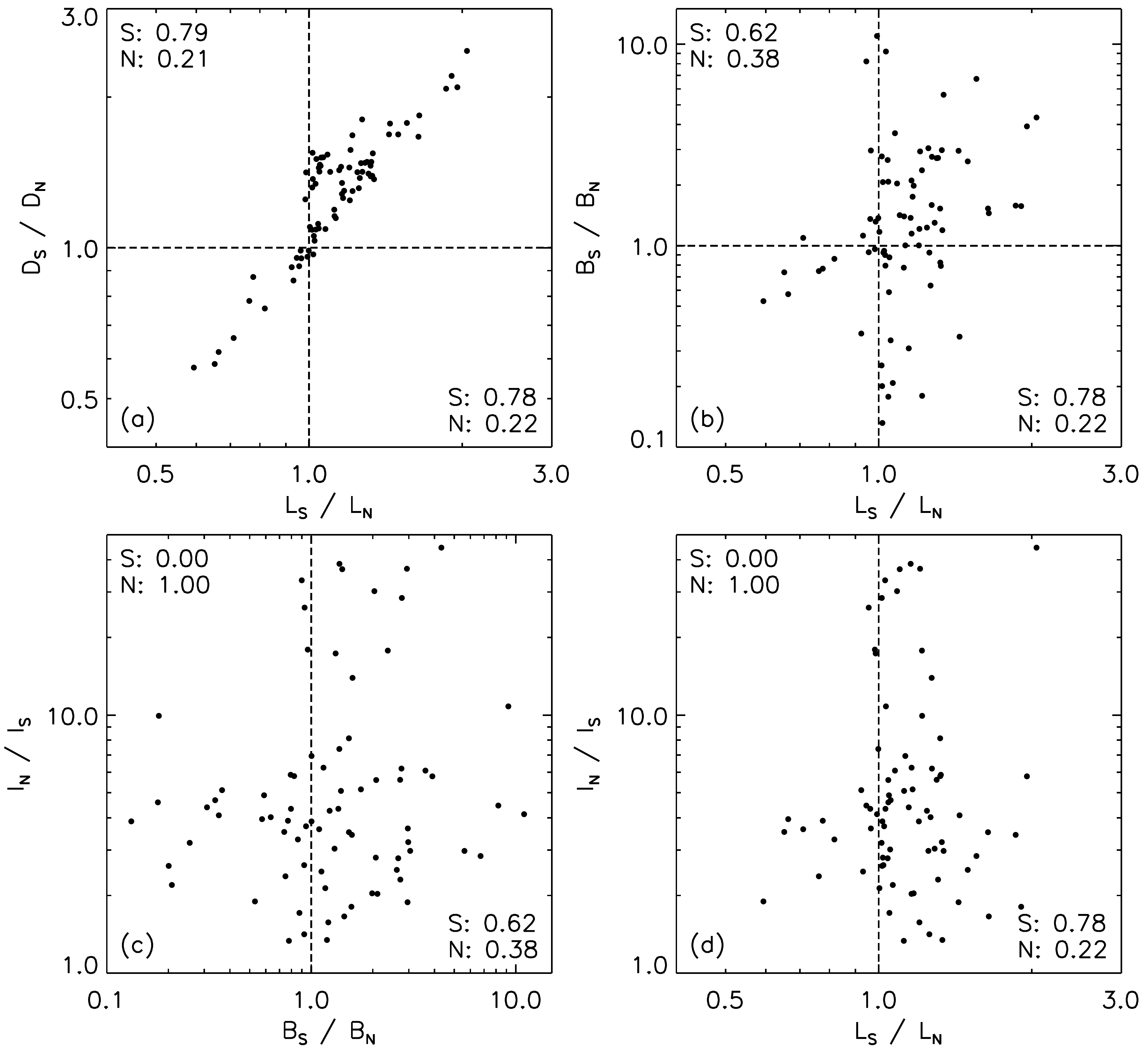}}
\caption{(\textbf{a}) Scatter diagram of the ratios between the de-projected length $L_{\rm S}/L_{\rm N}$ and the projected length $D_{\rm S}/D_{\rm N}$ from the LT to the FPs for each extrapolated field line. $L$ represents the integrated length from the LT to the FPs along each field line, while $D$ represents the distance between the projected point of the LT onto the XOY plane and the FPs. (\textbf{b}) Comparisons between the ratios of the integrated length $L_{\rm S}/L_{\rm N}$ from the LT to the FPs and the magnetic field strength $B_{\rm S}/B_{\rm N}$ at the FPs. Comparisons of the ratios of the HXR intensity $I_{\rm N}/I_{\rm S}$ with magnetic field strength $B_{\rm S}/B_{\rm N}$ at FPs (\textbf{c}) and length $L_{\rm S}/L_{\rm N}$ from the LT to the FPs (\textbf{d}). $B$ and $I$ represent the magnetic field strength and the HXR radiation at the FPs, respectively. The subscript `S' denotes the southern FP, while `N' denotes the northern FP. The value after the letter `S' (`N') in the corners indicates the percentage of the magnetic field lines where the physical quantity at the southern FP is larger (smaller) than that at the northern FP.}
\label{fig7}
\end{figure}

In addition, we conduct a statistical analysis on the relationship between the distance from LT to FPs, magnetic field strength $B$, and HXR intensity $I$ for each traced field line, as shown in Figure \ref{fig7}. In Panel a, we calculate the ratio of the de-projected length $L_{\rm S}/L_{\rm N}$ from the LT to the northern FT $L_{\rm N}$ and the southern FT $L_{\rm S}$, as well as the ratio of the distance $D_{\rm S}/D_{\rm N}$ from the projected points of the LT onto the XOY plane to the two FPs, respectively. The subscript `S' denotes the southern FP, while `N' denotes the northern FP. We find that 78\% of the traced magnetic field lines have $L_{\rm S} > L_{\rm N}$, while 22\% of the remains have $L_{\rm S} < L_{\rm N}$. Similarly, 79\% of magnetic field lines have $D_{\rm S} > D_{\rm N}$, whereas 21\% have $D_{\rm S} < D_{\rm N}$. The statistical relationship quantitatively demonstrates that the position of the flare LT is asymmetric. We then compare the relationship between the de-projected length ratio $L_{\rm S}/L_{\rm N}$ from LT to FPs and the ratio of the magnetic field strength $B_{\rm S}/B_{\rm N}$ at FPs, which is presented in Panel b. Based on the distribution of $B_{\rm S}/B_{\rm N}$, there is a slight difference in the magnetic field strength between the northern FPs and the southern FPs (62\% of magnetic field lines have $B_{\rm S} > B_{\rm N}$ at their FPs). The comparisons reveal that the asymmetry of the LT position is mainly attributed to the horn-shaped magnetic field configuration and is slightly influenced by the magnetic field strength at the FPs.

Furthermore, we compare the ratio of the HXR intensity $I_{\rm N}/I_{\rm S}$ at the FPs with the magnetic field strength $B_{\rm S}/B_{\rm N}$ and the loop length $L_{\rm S}/L_{\rm N}$, as shown in Figure \ref{fig7}c and d. Due to the uniform sampling of the northern FP for each magnetic field line within the contour of 30\% maximum HXR intensity, the HXR intensity ratios between the northern and southern FPs are always greater than 1, i.e., $I_{\rm N} > I_{\rm S}$. Similarly, one can see that there is a small difference in the magnetic field strength between the northern FP (38\% field lines of $B_{\rm S} < B_{\rm N}$) and the southern FP (62\% field lines of $B_{\rm S} > B_{\rm N}$), while 78\% of LTs are closer to the northern FP, $L_{\rm S} > L_{\rm N}$. Therefore, we conclude that the asymmetric distribution of the HXR radiation is mainly attributed to the asymmetry of the flare LT, and is also slightly influenced by the magnetic mirror effect.

\section{Conclusions and Discussions} \label{S-conclusions}

In this work, we investigate the asymmetry of the HXR radiation at two ribbons in a C4.4 flare by combining multi-view and multi-waveband observations from the ASO-S, SDO, and SolO spacecraft. The LDE observed on March 20, 2023, was caused by the eruption of a filament system, accompanied by a CME and an elongated CS. The light curves of both the SXR and HXR emissions present a consistent peak time, with no brightening observed in the 1600 {\AA} and 1700 {\AA} passbands, implying that thermal electrons dominate this event. The light curves of HXR, SDI {\hi} {\lya} and AIA {\heii} 304 {\AA} during the flare rising phase exhibit a consistent trend, indicating a common radiation origin primarily from flare ribbons formed in the middle-to-upper chromosphere and transition region.

Thanks to the 24-hour uninterrupted full-disk observations from Earth's view provided by the SDI on board the ASO-S, we can study the spatio-temporal features of the two ribbons in {\hi} {\lya} waveband. These observations, along with those in the {\heii} 304 {\AA}, reveal similar evolution of the two ribbons, presenting rapid separation towards the northern and southern directions. The pixel-by-pixel comparisons between the SDI and AIA images further quantitatively demonstrate the moderate correlation between {\hi} {\lya} and {\heii} 304 {\AA}. Meanwhile, their emissions may be closely associated with the heating process in the flare.

The spectral fitting of the HXR emission detected by STIX further confirms that the flare mainly releases energy through the process of thermal bremss-trahlung emission, implying that the radiations in both {\heii} 304 {\AA} and {\hi} {\lya} are produced by a thermal process. Additionally, the MEM\_GE algorithm is applied to reconstruct the HXR maps observed by STIX. We find that the HXR radiation presents an asymmetric distribution, with the majority of the emission concentrated at the northern ribbon.

Using the method of differential emission measure \citep[DEM,][]{Cheung2015ApJ,Su2018ApJ}, we estimate the electron density $n_{\rm e}$ inside the flare loops. The total emission measure EM and the average electron number density are deduced via,
\begin{eqnarray}
{\rm EM} &=& \int_{T_{\rm min}}^{T_{\rm max}}{{\rm DEM}(T)dT}, \\
n_{\rm e} &=& \sqrt{\frac{\rm EM}{l}}\ ({\rm cm}^{-3}).
\end{eqnarray}
where $l$ represents the effective depth along LoS. We further assume that the flare loops have similar extents in depth and width, and we estimate $l$ to be $\approx2.3\times10^{8}\ {\rm cm}$ by measuring the loop width observed by AIA 94 {\AA}. The value of EM can be directly derived from the DEM result. Therefore, we determine the density $n_{\rm e}$ is $\approx1.0\times10^{10}\ {\rm cm^{-3}}$, indicating that the flare loop is relatively dense.

By combining the NLFFF extrapolation with observations, we explore the magnetic configuration and physical parameters within this active region. This approach allows us to understand the magnetic field structure and derive important physical properties within the two ribbons. The characteristic loop length $L$, derived from the traced field lines, is $\approx5.9\times10^9\ {\rm cm}$. Based on the magnetic topology, we further demonstrate that the asymmetric HXR emission is produced by the asymmetric position of the flare LT and the magnetic mirror effect. The statistical parameters of the extrapolated field lines indicate that the de-projected length from the LT to the northern FP is shorter than that to the southern FP, and the magnetic field strength at the southern FP is stronger than that at the northern FP, which promotes thermal electrons to collide with the plasma inside the dense flare loop and easily reach the northern ribbon. At the same time, considering the physical conditions such as the lower height, $\approx32\ {\rm Mm}$, of the LT above the photosphere, as well as the denser plasma inside the flare loops, $\approx1.0\times10^{10}\ {\rm cm^{-3}}$, more thermal electrons are more easily able to reach the northern FP during the collision process with the surrounding plasma, resulting in a stronger HXR emission compared to the southern FP.

Our work provides a novel approach for investigating the asymmetric HXR radiation in the two-ribbon flare dominated by over 95\% thermal radiation. This approach involves combining the NLFFF extrapolation with observations to understand the magnetic field structure and physical properties within the flare region. Unfortunately, due to no obvious HXR emissions above $\approx18$ keV in this flare, the HXI on board the ASO-S did not receive sufficient counts for data analysis. Future research endeavors will statistically analyze the asymmetric events of the HXR emission at FPs using the HXR data observed by the HXI and combine them with the unique {\hi} {\lya} waveband full-disk observations from the SDI to conduct more detailed studies of the asymmetric HXR radiation.

%%%%%%%%%%%%%%%%%%%%%%%%%%%%%%%%%%%%%%%%%%%%%%%%%%%%%%%%%%%%%%%%%%%%%%%%%%%

\begin{acks}

We sincerely thank the anonymous referee for providing valuable suggestions that helped us improve the quality of the manuscript. We thank Shangbin Yang and Zhentong Li for their helpful discussions on the NLFFF model and STIX data processing. The authors thank the teams of SolO/STIX, SDO, GOES, CHASE for their open-data use policy. The ASO-S mission is supported by the Strategic Priority Research Program on Space Science, Chinese Academy of Sciences. SolO is a space mission of international collaboration between ESA and NASA, operated by ESA. The STIX instrument is an international collaboration between Switzerland, Poland, France, Czech Republic, Germany, Austria, Ireland, and Italy. SDO is a mission of NASA's Living with a Star (LWS) Program. The CHASE mission is supported by China National Space Administration.

\end{acks}

\begin{authorcontribution}

G.L. Shi wrote the main manuscript, analyzed the data, and generated figures. L. Feng discovered the asymmetric HXR radiation event, conceived the study, and revised the manuscript. J. Chen, B.L. Ying, S.T. Li, and Q. Li discussed results and provided helpful suggestions on this work. W.Q. Gan is PI of the ASO-S. H. Li and L. Feng are PI and Co-PI of the LST, respectively. K.F. Ji provided an alignment code for the AIA and SDI images. Y. Li, Y. Huang, Y.P. Li, J.W. Li, J. Zhao, L. Lu, J.C. Xue, P. Zhang, D.C. Song, Z.Y. Tian, Y.N. Su, Q.M. Zhang, Y.Y. Ge, J.H. Shan, Y. Zhou, J. Tian, G. Li, X.F. Liu, Z.C. Jing, and S.J. Lei contributed to the in-orbit testing, pipeline, and release of LST data. All authors reviewed the manuscript.

\end{authorcontribution}

\begin{fundinginformation}

This work is supported by the National Key R\&D Program of China 2022YFF-0503003 (2022YFF0503000), Strategic Priority Research Program of the Chinese Academy of Sciences, Grant No. XDB0560000, NSFC (grant Nos. 11973012, 11921003, 12103090, 12203102, 12233012, 12373115), the mobility program (M-0068) of the Sino-German Science Center.

\end{fundinginformation}

\begin{dataavailability}

The ASO-S data before Apr. 1, 2023 are currently being tested and not publicly available, but can be obtained from the corresponding author with reasonable request. The ASO-S data after Apr. 1, 2023 are publicly available at \url{http://aso-s.pmo.ac.cn/sodc/dataArchive.jsp}. The AIA\&HMI data are downloaded from the Joint Science Operations Center (JSOC) at \url{http://jsoc.stanford.edu}. The STIX data are publicly available at \url{https://datacenter.stix.i4ds.net/view/list/fits}. The GOES data are obtained from the National Oceanic and Atmospheric Administration (NOAA) at \url{https://www.ngdc.noaa.gov/}. The CHASE data are accessible through the solar science data center of Nanjing university at \url{https://ssdc.nju.edu.cn/NdchaseSatellite}.

\end{dataavailability}

\begin{ethics}
\begin{conflict}
The authors declare no competing interests.
\end{conflict}
\end{ethics}

\bibliographystyle{spr-mp-sola}
\bibliography{glshi_bibliography}

\end{document}